\documentstyle[aps]{revtex}
\tighten
\begin{document}
%\draft
%======================================%
%<<<<<<<<<<<< TITLE PAGE >>>>>>>>>>>>>>%
%======================================%
\preprint{HUTP-03/A066,hep-th/0310xxx}
\title{Scaling solution, radion stabilization, and initial condition for
brane-world cosmology} 
\author{Shinji Mukohyama}
\address{
Department of Physics, Harvard University\\
Cambridge, MA, 02138, USA
}
\author{Alan Coley}
\address{
Department of Mathematics \& Statistics,
Dalhousie University\\
Halifax, Nova Scotia, Canada  \enskip  B3H 3J5
}
\date{\today}

\maketitle

%======================================%
%<<<<<<<<<<<<< ABSTRACT >>>>>>>>>>>>>>>% 
%======================================%
\begin{abstract} 
 We propose a new, self-consistent and dynamical scenario which gives
 rise to well-defined initial conditions for five-dimensional
 brane-world cosmologies with radion stabilization. At high energies,
 the five-dimensional effective theory is assumed to have a scale
 invariance so that it admits an expanding scaling solution as a future
 attractor. The system automatically approaches the scaling solution
 and, hence, the initial condition for the subsequent low-energy brane
 cosmology is set by the scaling solution. At low energies, the scale
 invariance is broken and a radion stabilization mechanism drives the
 dynamics of the brane-world system. We present an exact, analytic
 scaling solution for a class of scale-invariant effective theories of
 five-dimensional brane-world models which includes the five-dimensional
 reduction of the Horava-Witten theory, and provide convincing evidence
 that the scaling solution is a future attractor. 
 \hfill\mbox{[HUTP-03/A066]}
\end{abstract}

%======================================%
%<<<<<<<<<<<< SECTION I  >>>>>>>>>>>>>>%
%======================================%
\section{Introduction}

% RS2 brane-world 

Phenomenological models in which our universe is a four-dimensional
($4$d) hypersurface, or a $3$-brane, embedded in a higher-dimensional
spacetime is currently of great interest. In the brane-world scenario,
ordinary matter fields are confined to the brane, while the
gravitational field can propagate in the extra dimensions (i.e., in the
`bulk')~\cite{earlier-works,ADD,AADD,RS1,RS2}. In particular, Randall
and Sundrum proposed a self-consistent brane-world (RS2) scenario with
$5$d anti-de Sitter (AdS) bulk spacetime and showed that $4$d gravity is
also localized  on the brane if the brane tension is positive and 
fine-tuned~\cite{RS2}. Many aspects of the Randall-Sundrum scenarios
have been investigated, including weak
gravity~\cite{Garriga-Tanaka,SSM,GKR}, the effective Einstein
equation~\cite{SMS,Maartens},
cosmology~\cite{CGKT,CGS,FTW,BDEL,Mukohyama2000a,Kraus,Ida,MSM}, black 
holes~\cite{CHR,EHM,SS,KTN,BH-Holography} and so on.

% RS2 brane-world cosmology

In the brane-world scenario, the evolution of our universe can be
considered in two ways. On the one hand, it is the expansion of the 
induced geometry on our brane. Note that ordinary matter fields are
assumed to be confined on the brane and, thus, propagate in the induced 
geometry. This picture is not very different from what we are accustomed
to in the standard $4$d cosmology. Indeed, the generalized Friedmann
equation for homogeneous, isotropic brane universe in the RS2 scenario
reduces to the standard Friedmann equation at low energy with
corrections~\cite{CGKT,CGS,FTW,BDEL,Mukohyama2000a}. In this picture the
bulk geometry also evolves~\footnote{Here, the bulk geometry is supposed
to be seen in the Gaussian normal coordinate based on the world-volume
of our brane.}, corresponding to the expansion of our universe. On the
other hand, the same evolution of our universe can be seen in a
different way as a brane motion in a higher dimensional spacetime. In
this picture the generalized Friedmann equation is understood as the
equation of motion for the brane moving in the bulk spacetime. In the
RS2 scenario this picture is remarkably simple for the homogeneous,
isotropic brane universe. Indeed, the bulk spacetime is always locally
AdS-Schwarzschild spacetime, irrespective of the brane
motion~\cite{Kraus,Ida,MSM}. In other words, the brane motion does not
generate waves in the bulk. Unfortunately, in the more realistic
scenarios which we shall consider in this paper this remarkable property
is not satisfied. Nonetheless, it is still useful to keep in mind that
there are two different ways to describe the same evolution of the
brane-world cosmology: the expansion of the induced geometry and the
brane motion in the bulk.

% RS1 brane-world

Prior to the RS2 scenario, Randall-Sundrum had proposed a similar but 
different scenario called the RS1 scenario~\cite{RS1}. This scenario
involves two branes with fine-tuned tensions, one positive and another
negative, and aims to solve the gauge hierarchy problem. In order to
obtain an appropriate hierarchy between the Planck scale and the
electroweak scale, the distance between the two branes must be set to
about $50$ times the bulk curvature scale. This itself is not a severe 
fine-tuning nor a problem at all, but it would be more satisfactory if
this value could be dynamically realized. In the RS1 scenario,
unfortunately, from the $4$d brane viewpoint the inter-brane distance
represents a massless scalar degree of freedom and, thus, is
arbitrary. Moreover, the existence of this massless scalar mode, called 
the {\it radion}, causes a more serious problem: the low energy $4$d
gravity on our brane is not general relativity but a Brans-Dicke type
theory with a too small Brans-Dicke parameter to be consistent with
gravitational experiments~\cite{Garriga-Tanaka}. Here, the radion plays
the roll of the Brans-Dicke scalar.

% Radion stabilization and 4d Einstein gravity

The problem of too large a deviation from the Einstein theory can be
fixed if the radion obtains a non-zero mass. With a non-vanishing radion
mass, the radion is not excited and becomes irrelevant at energies much
lower than the radion mass, so that the $4$d Einstein gravity is
restored at low energy. This is called {\it radion stabilization}. 
Goldberger and Wise proposed a mechanism to stabilize the radion by
introducing a scalar field in the bulk~\cite{GW}. This scalar field has
a potential in the bulk (bulk potential) and couples to branes via its
potentials localized on the branes (brane potentials). The introduction
of scalar field(s) in the bulk is, of course, favorable from the
viewpoint of string theory since a compactification from $10$ or $11$
dimensions down to $5$ dimensions in general introduces many $5$d scalar
fields. In this sense, the generalized RS1 scenarios with radion
stabilization seem more realistic than the RS2 model. With the
Goldberger-Wise mechanism, $4$d Einstein gravity is indeed shown to be
recovered on our 
brane~\cite{Tanaka-Montes,CGK,Kudoh-Tanaka,Mukohyama-Kofman,Mukohyama-HDterm}~\footnote{For
an apparent conflict with the picture in ref.~\cite{MW}, and 
its resolution, see ref.~\cite{Mukohyama-Kofman}.}.
The radion stabilization also helps recovery of the standard Friedmann
equation at low energy~\cite{CGR,KKOP}. 

% Brane cosmology with radion stabilization

The introduction of the bulk scalar complicates the brane-world scenario
in (at least) two different ways. First, the gauge hierarchy problem is
now entangled with the cosmological constant problem. The essential
reason for this is that both the $4$d cosmological constant and the
stabilized value of the inter-brane distance depend on the bulk
potential and the brane potentials of the scalar field in a non-trivial
way.

Second, the dynamics of the system becomes much more complicated. The
bulk geometry is no longer simple if a brane is moving; i.e., if our
$4$d universe is expanding. A general brane motion generates waves of
the bulk scalar field and makes the bulk geometry very complicated. This
makes it extremely difficult to analyze the general dynamics of
brane-world cosmology analytically. Recent development of
BraneCode~\cite{branecode} has made it possible to analyze the dynamics
numerically, but there remains the problem of initial
conditions. Namely, the system is so rich that it is not a priori
trivial to choose physically relevant initial conditions for the
numerical study.

The purpose of this paper is to shed light on this second point in a
particular way. In the rest of this paper, our main concern will be 
the dynamics of the brane-world system with radion stabilization. In
particular, we shall propose a new dynamical scenario to give an initial
condition for the brane-world cosmology with radion stabilization.

\section{Physical scenario} 
\label{sec:scenario}

Suppose that the $5$d theory has an approximate scale invariance at
relatively high energy (but still much lower than the Planck scale) so
that the $5$d effective action allows a scaling solution. For example,
we might imagine that the bulk potential $V(\phi)$ and the brane
potentials $\lambda_{\pm}(\phi_{\pm})$ have the form 
%============< EQUATION >==============%
%
\begin{equation}
 V(\phi) = V_0e^{-2\alpha\kappa_5\phi}, \quad
  \lambda_{\pm}(\phi_{\pm}) = \lambda_0^{\pm}
  \left[e^{-\alpha\kappa_5\phi_{\pm}} + e^{\beta_{\pm}\kappa_5\phi_{\pm}}
  \right], \label{eqn:example-pot}
\end{equation}
%======================================%
where $\phi_{\pm}$ is the pullback of $\phi$ on each brane and
$\alpha+\beta_{\pm}\ne 0$. We assume that
$-(\alpha+\beta_{\pm})\kappa_5\phi_{\pm}\gg 1$ initially so that the
term $e^{\beta_{\pm}\kappa_5\phi_{\pm}}$ can be neglected and the system
has the scaling invariance.

It is worth mentioning here that potentials of the form
(\ref{eqn:example-pot}) arise in many theories of the fundamental
interactions. For example, compactification of higher-dimensional
theories, including superstring theories, gives rise to exponential
terms in the effective potential. Terms with different exponents come
from different physical effects, eg., higher-dimensional cosmological
constant, curvature of compact manifold, anti-symmetric field flux,
Casimir effects, etc. We shall provide more motivations for exponential
potentials in subsection~\ref{subsec:action}.

Since the inter-brane distance has the dimension of length, we
expect that it should be proportional to time and, thus, expanding for
the scaling solution. In this paper, we shall find such a scaling
solution and present convincing evidence that it is a future
attractor. For the moment, we just assume that the scaling solution is a
future attractor. In this case, the system should approach it
dynamically. Subsequently, since the Hubble expansion rate decays as 
$1/t$ and the size of the extra-dimension increases as $t$ for the
scaling solution, the system should enter a lower and lower energy
regime. Hence, the scale invariance is no more than an approximate
symmetry. In the above example (\ref{eqn:example-pot}), when
$-(\alpha+\beta_{\pm})\kappa_5\phi_{\pm}$ becomes of order unity, the
term $e^{\beta_{\pm}\kappa_5\phi_{\pm}}$ in the brane potentials cannot
be neglected any more and the approximate scaling invariance will be
broken.

Depending on the form of the effective action relevant to the lower
energy regime, the inter-brane distance, called the radion, may play the
role of an inflaton. Hence, we will have a $4$d inflation on our brane
driven by the radion. When the radion-induced inflation ends, the radion
is stabilized and the fields confined on our brane can be reheated due
to the oscillatory behavior of the system around the stabilized
configuration. Subsequently, the conventional standard cosmology is
realized. Since the radion is eventually stabilized, Einstein weak
gravity is restored at low
energy~\cite{Tanaka-Montes,Kudoh-Tanaka,Mukohyama-Kofman,Mukohyama-HDterm}.

Hence, in this hypothetical scenario, the role of the scaling solution
is to provide the initial condition for the radion-induced
inflation and the subsequent evolution. To our knowledge there has been 
no well-defined scenario to provide an initial condition for the
brane-world cosmology, except perhaps for the creation-from-nothing
scenario~\cite{Garriga-Sasaki}. If the scaling solution is a future
attractor then the classical dynamics will automatically drive the
system to the well-defined initial condition for the following evolution
of the brane world cosmology.

In the remainder of this paper we shall find an exact, analytic scaling
solution and present convincing evidence for its attractor behavior.

%======================================%
%<<<<<<<<<<<< SECTION III >>>>>>>>>>>>>%
%======================================%
\section{Basic equations}

In this section we present the basic equations for general
potentials. In the following sections we apply these equations to a
specific model with exponential potentials, motivated by string-inspired
phenomenology.

In the $5$d bulk we consider Einstein gravity and a scalar
field:
%============< EQUATION >==============%
%
\begin{equation}
 I_5 = \int dx^5 \sqrt{-g}
  \left[ \frac{R}{2\kappa_5^2} 
   - \frac{1}{2}\partial^M\phi\partial_M\phi
   - V(\phi)\right]. 
\end{equation}
%======================================%
Hence, the $5$d Einstein equation is 
%============< EQUATION >==============%
%
\begin{eqnarray}
 G_{MN} & = & \kappa^2 T_{MN}, \nonumber\\
 T_{MN} & = & \partial_M\phi\partial_N\phi
  -\left[\frac{1}{2}\partial^L\phi\partial_L\phi
   + V(\phi)\right]g_{MN}. 
\end{eqnarray}
%======================================%
The field equation for $\phi$,
%============< EQUATION >==============%
%
\begin{equation}
 \nabla^2\phi - V'(\phi) = 0,
\end{equation}
%======================================%
where $\nabla$ is the covariant derivative compatible with the metric
$g_{MN}$, follows automatically from the Einstein equation because of
the Bianch identity $\nabla^{\mu}G_{\mu\nu}=0$.

We assume that the extra dimension has the topology of $Z_2/S^1$ and
consider two end-of-the-world branes on the two fixed points. 
The tension $\lambda_{\pm}$ of each brane in general depends on the
pullback $\phi_\pm$ of the scalar field $\phi$. Hence, the brane action
is of the form 
%============< EQUATION >==============%
%
\begin{equation}
 I_4 =   -\int dx_-^4 \sqrt{-q_-}\lambda_-(\phi_-)
  -\int dx_+^4 \sqrt{-q_+}\lambda_+(\phi_+),
\end{equation}
%======================================%
where $x_{\pm}$ represents $4$d coordinates on each brane and
$q_{\pm}$ is the determinant of the induced metric.

Note that we need to add the Gibbons-Hawking boundary term to the
action, depending on the precise definition of $I_5$. If $I_5$ 
includes the integration over the thin layers corresponding to the
branes then the Gibbons-Hawking term must not be added since it appears
automatically as we integrate the Einstein-Hilbert term over the thin
layers~\cite{Mukohyama-action}. On the other hand, if $I_5$ does not 
include the integration over the thin layers then the Gibbons-Hawking
boundary term must be added. In all cases, including the variation of
the metric, the position of the hypersurface and the scalar field, the 
variational principle gives the correct set of equations of motion: the
Einstein equation and the field equation of the scalar field off the 
branes, Israel's junction condition and the matching condition for the
scalar field~\cite{Mukohyama-action}.

\subsection{Bulk equation}

For simplicity we assume that the geometry on each brane is everywhere 
and for all times described by the $4$d flat FRW universe and
that the two branes are parallel to each other. In this case, it is
natural to expect that the $5$d bulk spacetime between the
two branes also has the same symmetry. Hence, we assume that the metric
and the scalar field have the following forms:
%============< EQUATION >==============%
%
\begin{eqnarray}
 ds^2 & = & -n(t,y)^2dt^2 + a(t,y)^2\delta_{ij}dx^idx^j + b(t,y)^2dy^2, 
  \nonumber\\
 \phi & = & \phi(t,y),
\end{eqnarray}
%======================================%
where $i=1,2,3$; $y$ represents the extra dimension; and the
world-volume of each brane is expressed as
%============< EQUATION >==============%
%
\begin{equation}
 y = Y_{\pm}(t), \quad (Y_-(t)<Y_+(t)).
\end{equation}
%======================================%
For this ansatz, we obtain
%============< EQUATION >==============%
%
\begin{eqnarray}
 \frac{G_{tt}}{n^2} & = & 
  \frac{3}{n^2}\left[ \left(\frac{\dot{a}}{a}\right)^2
   +\frac{\dot{a}}{a}\frac{\dot{b}}{b} \right]
   +\frac{3}{b^2}\left[
   -\frac{a''}{a} -\left(\frac{a'}{a}\right)^2
   +\frac{a'}{a}\frac{b'}{b} \right], \nonumber\\
 G_{ty} & = & 
  3\left[ -\frac{\dot{a}'}{a} + \frac{\dot{a}}{a}\frac{n'}{n}
    + \frac{a'}{a}\frac{\dot{b}}{b} \right], \nonumber\\
 \frac{G_{yy}}{b^2} & = & \frac{3}{n^2}
  \left[ -\frac{\ddot{a}}{a} -\left(\frac{\dot{a}}{a}\right)^2
   + \frac{\dot{a}}{a}\frac{\dot{n}}{n}\right]
  +\frac{3}{b^2}\left[
     \left(\frac{a'}{a}\right)^2 + \frac{a'}{a}\frac{n'}{n}
    \right],\nonumber\\
 \frac{G_{ij}}{a^2} & = & 
  \frac{\delta_{ij}}{n^2}
   \left[ -2\frac{\ddot{a}}{a} - \left(\frac{\dot{a}}{a}\right)^2
    +2\frac{\dot{a}}{a}\frac{\dot{n}}{n}
    -2\frac{\dot{a}}{a}\frac{\dot{b}}{b}
    +\frac{\dot{b}}{b}\frac{\dot{n}}{n}
    -\frac{\ddot{b}}{b} \right]
   +\frac{\delta_{ij}}{b^2}
   \left[ 2\frac{a''}{a} + \left(\frac{a'}{a}\right)^2
    + 2\frac{a'}{a}\frac{n'}{n} - 2\frac{a'}{a}\frac{b'}{b}
    -\frac{b'}{b}\frac{n'}{n} + \frac{n''}{n}\right], \nonumber\\ 
 G_{ti} & = & G_{yi} = 0,
\end{eqnarray}
%======================================%
and
%============< EQUATION >==============%
%
\begin{eqnarray}
 \frac{T_{tt}}{n^2} & = & 
  \frac{1}{2}\left(\frac{\dot{\phi}}{n}\right)^2
  + \frac{1}{2}\left(\frac{\phi'}{b}\right)^2  + V, \nonumber\\
 T_{ty} & = & \dot{\phi}\phi', \nonumber\\
 \frac{T_{yy}}{b^2} & = & 
  \frac{1}{2}\left(\frac{\dot{\phi}}{n}\right)^2
  + \frac{1}{2}\left(\frac{\phi'}{b}\right)^2  - V, \nonumber\\
 \frac{T_{ij}}{a^2} & = & 
  \frac{1}{2}\left(\frac{\dot{\phi}}{n}\right)^2
  - \frac{1}{2}\left(\frac{\phi'}{b}\right)^2  - V, \nonumber\\
 T_{ti} & = & T_{yi} = 0,
\end{eqnarray}
%======================================%
where a dot and a prime denote derivatives with respect to $t$ and $y$,
respectively.

\subsection{Boundary condition}

On each brane at $y=Y_{\pm}(t)$, the induced metric $q_{\pm\mu\nu}$, the
extrinsic curvature $K_{\pm\mu\nu}$ and the pullback $\phi_{\pm}$ of
the scalar field play important roles. They are given by  
%============< EQUATION >==============%
%
\begin{eqnarray}
 q_{\pm tt}
  & = & -\left[n^2-b^2\dot{Y}_{\pm}^2\right], \nonumber\\
 q_{\pm ij}
  & = & a^2\delta_{ij}, \nonumber\\
 {\cal K}_{\pm} & \equiv & K^{\ t}_{\pm t} = 
  \left[1-\left(\frac{b\dot{Y}_{\pm}}{n}\right)^2\right]^{-\frac{3}{2}}
  \cdot
  \left[\frac{b\ddot{Y}_{\pm}}{n^2} 
  - \left(\frac{b\dot{Y}_{\pm}}{n}\right)^3\frac{\dot{b}}{bn} 
  + \left(\frac{b'}{b^2}-\frac{2n'}{bn}\right)
  \left(\frac{b\dot{Y}_{\pm}}{n}\right)^2 
  + \left(\frac{2\dot{b}}{bn}-\frac{\dot{n}}{n^2}\right)
  \frac{b\dot{Y}_{\pm}}{n}
  +\frac{n'}{bn}\right],   \nonumber\\
 \bar{\cal K}_{\pm}\delta^i_j & \equiv & K^{\ i}_{\pm j} = 
  \delta^i_j\cdot
  \left[1-\left(\frac{b\dot{Y}_{\pm}}{n}\right)^2\right]^{-\frac{1}{2}}
  \cdot
  \left[\frac{a'}{ab} + \frac{\dot{a}}{an}\frac{b\dot{Y}_{\pm}}{n}\right]. 
  \nonumber\\
 \phi_{\pm} & = & \phi(t,Y_{\pm}), \nonumber\\
 \partial_{\perp}\phi_{\pm} & \equiv & 
  \left.n^{\mu}\partial_{\mu}\phi\right|_{y=Y_{\pm}(t)} = 
  \left[1-\left(\frac{b\dot{Y}_{\pm}}{n}\right)^2\right]^{-\frac{1}{2}}
  \cdot
  \left[\frac{b\dot{Y}_{\pm}}{n}
   \frac{\dot{\phi}}{n}+\frac{\phi'}{b}\right]. 
\end{eqnarray}
%======================================%
Having the $Z_2$ symmetry, the Israel junction condition and the scalar
field matching condition are written as
%============< EQUATION >==============%
%
\begin{eqnarray}
 {\cal K}_{\pm} & = & \pm\frac{\kappa_5^2}{6}\lambda_{\pm}(\phi_{\pm}),
  \nonumber\\ 
 \bar{\cal K}_{\pm} & = & \pm\frac{\kappa_5^2}{6}\lambda_{\pm}(\phi_{\pm}), 
  \nonumber\\
 \partial_{\perp}\phi_{\pm} & = & 
  \mp\frac{1}{2}\partial_{\phi_{\pm}}\lambda_{\pm}(\phi_{\pm}). 
  \label{eqn:general-boundary-cond}
 \end{eqnarray}
%======================================%
When $Y_{\pm}(t)$ are constants $y_{\pm}$, these conditions reduce to
the following simple conditions 
%============< EQUATION >==============%
%
\begin{eqnarray}
 \left.\frac{n'}{bn}\right|_{y=y_{\pm}} & = &
  \left.\frac{a'}{ba}\right|_{y=y_{\pm}}
  = \pm\frac{\kappa_5^2}{6}\lambda_{\pm}(\phi_{\pm}),
  \label{eqn:junction-cond}\\
 \left.\frac{\phi'}{b}\right|_{y=y_{\pm}} & = &
  \mp\frac{1}{2}\partial_{\phi_{\pm}}\lambda_{\pm}(\phi_{\pm}).
  \label{eqn:matching-cond}
\end{eqnarray}
%======================================%

%======================================%
%<<<<<<<<<<<< SECTION IV >>>>>>>>>>>>>>%
%======================================%

\section{Scaling solution}

In this section we shall seek a scaling solution for a specific model
with exponential potentials. For a special choice of parameters, the
model reduces to the $5$d reduction of the Horava-Witten
theory~\cite{LOSW}. On the other hand, the solution we shall find does
not appear to reduce to the known flat FRW solution in the Horava-Witten
theory\cite{LOW}. (For closed and open FRW solutions in Horava-Witten
theory, see ref.~\cite{Reall}.)

\subsection{Action}
\label{subsec:action}

Exponential potentials of the form
%============< EQUATION >==============%
%
\begin{equation}
 V= V_0 e^{-2\alpha\kappa\phi}
\end{equation}
%======================================%
for (dilaton coupling) constant $\alpha$, arise in many theories of the
fundamental interactions including superstring and higher-dimensional
theories. Typically, `realistic' supergravity theories predict steep
exponential potentials \cite{Green} (i.e., $2\alpha^2>1$). The effective
action of the Horava-Witten theory  has $\alpha^2=2$ (see below). A
smaller {\em effective} $\alpha$  can arise in assisted theories which
contain more than one scalar field \cite{Liddle}. In addition, for many
compactifications of higher-dimensional theories there is a consistent
truncation to a single scalar field $\phi$ and positive potentials of
this form which arises via generation by non-zero flux of antisymmetric
tensor fields (the $T^7$ compactification of $11$d supergravity with 
non-vanishing 4-form field strength yields an exponential potential with
$\alpha=\sqrt{7}$; in general, flux compactifications seem to yield
$\alpha\ge\sqrt{3}$ \cite{ANT}), and hyperbolic compactifications  in
the context of String/M-theory \cite{BDLPS,russo} (the compactification
of $11$d supergravity on a $7$d compact hyperbolic  
space has $\alpha= 3/\sqrt{7}$; in general, hyperbolic compactifications
seem to lead to $1<\alpha<\sqrt{3}$ \cite{townsend}). A comprehensive
qualitative analysis of scalar field cosmological models with an
exponential potential has been presented \cite{qual}.

In this paper we consider a brane-world analogue of such models with
exponential potentials. To motivate our model with exponential
potentials, let us briefly review the $5$d reduction of the
Horava-Witten theory. The $5$d effective action for the Horava-Witten
theory is~\cite{LOSW} 
%============< EQUATION >==============%
%
\begin{equation}
 I_{HW} = \frac{1}{2\kappa_5^2}\int dx^5\sqrt{-g}
  \left[R-\frac{\partial^M\Phi\partial_M\Phi}{2\Phi^2}
   -\frac{\tilde{\alpha}^2}{3\Phi^2}\right]
  + \frac{\sqrt{2}}{\kappa_5^2}
  \int dx_-^4\sqrt{-q_-}\frac{\tilde{\alpha}}{\Phi_-}
  - \frac{\sqrt{2}}{\kappa_5^2}
  \int dx_+^4\sqrt{-q_+}\frac{\tilde{\alpha}}{\Phi_+},
\end{equation}
%======================================%
where $\Phi_{\pm}$ is the pullback of the bulk scalar field $\Phi$ onto
each brane. If we introduce a canonically normalized scalar field $\phi$
by 
%============< EQUATION >==============%
%
\begin{equation}
 \sqrt{2}\kappa_5\phi \equiv 
  \ln\left(\frac{\Phi}{\sqrt{2}\tilde{\alpha}\kappa_5^{2/3}}\right), 
\end{equation}
%======================================%
this action is reduces to 
%============< EQUATION >==============%
%
\begin{equation}
 I_{HW} = \int dx^5\sqrt{-g}
  \left[\frac{1}{2\kappa_5^2}R
   -\frac{1}{2}\partial^M\phi\partial_M\phi
   -\frac{e^{-2\sqrt{2}\kappa_5\phi}}{12\kappa_5^{10/3}}\right]
  + \int dx_-^4\sqrt{-q_-}
  \frac{e^{-\sqrt{2}\kappa_5\phi_-}}{\kappa_5^{8/3}}
  - \int dx_+^4\sqrt{-q_+}
  \frac{e^{-\sqrt{2}\kappa_5\phi_+}}{\kappa_5^{8/3}},
\end{equation}
%======================================%
where $\phi_{\pm}$ is the pullback of the scalar field $\phi$ on each
brane.

The model we shall consider is
%============< EQUATION >==============%
%
\begin{equation}
 I = \int dx^5\sqrt{-g}
  \left[\frac{1}{2\kappa_5^2}R
   -\frac{1}{2}\partial^M\phi\partial_M\phi
   -V_0e^{-2\alpha\kappa_5\phi}\right]
  - \int dx_-^4\sqrt{-q_-}\lambda^-_0e^{-\alpha\kappa_5\phi_-}
  - \int dx_+^4\sqrt{-q_+}\lambda^+_0e^{-\alpha\kappa_5\phi_+},
  \label{eqn:scale-invariant-action}
\end{equation}
%======================================%
where $\alpha$, $V_0$ and $\lambda^{\pm}_0$ are constants. This action
includes the $5$d effective action of the Horava-Witten
theory as a special case ($\alpha = \sqrt{2}$).

\subsection{Ansatz}

We would like to seek a special solution which corresponds to an
`equilibrium state', which represents a future attractor for the
model. In a general situation where branes move arbitrarily, the motion
of branes produces waves of the scalar field $\phi$ in the bulk. The
waves in the bulk interact with both branes in the sense that they can
come and go between two end-of-the-world branes. Hence, the general
situation is very complicated, and non-local from the viewpoint of a
$4$d observer on a brane~\cite{Mukohyama-nonlocal}. On the
other hand, we would like to seek a special solution in which waves
emitted from one brane and those from another brane are in equilibrium
so that there is effectively no scalar wave in the bulk. Hence, what we
would like to seek is a very special situation in which the non-local
effects due to scalar waves are completely suppressed.

In this kind of 'equilibrium' situation without waves in the bulk,
nothing should propagate from one brane to another and, thus, we
expect that all physically meaningful functions of $t$ and $y$ should be
a product of a function of $t$ and a function of $y$. In particular, in
this case the time dependence of $n$ can be removed by a coordinate
choice. 
Moreover, from our experience in $4$d scalar field
cosmology with an exponential potential, we expect that the
'equilibrium' situation should correspond to a power-law expansion on
each brane, $a(t,y_{\pm})\propto t^p$ \cite{qual}. Actually, in
$4$d cosmology with an exponential potential, a power-low 
expansion is an attractor of the system. One of the essential physical
reasons for this is that all physically relevant, dimensionful
quantities scale in tandem, according to dimensionality. For example,
$H^2\propto \kappa_4^2V\propto\kappa_4^2\rho_{kin}\propto 1/t^2$, where
$H$ is the Hubble expansion rate, $\kappa_4$ is the $4$d
gravitational coupling, $V$ is the exponential potential and
$\rho_{kin}$ is the kinetic energy of the scalar field. 
In our model, besides the Hubble expansion rate on each brane,
the inter-brane distance is another physically relevant, dimensionful
quantity. Since it has the dimension of length, we expect it should also
be proportional to $t$ in the 'equilibrium' situation. Thus, not only
the $4$d universes on branes but also the inter-brane distance should be
expanding.

For the reasons explained above, we consider the following ansatz. 
%============< EQUATION >==============%
%
\begin{eqnarray}
 ds^2 & = & n(y)^2\left(-dt^2 + t^{2p}\delta_{ij}dx^idx^j + t^2dy^2\right), 
  \nonumber\\
 \kappa_5^2|V_0|e^{-2\alpha\kappa_5\phi} & = & 
  \frac{e^{-2\psi(y)}}{l^2t^2},\nonumber\\
 Y_{\pm}(t) & = & y_{\pm},
\end{eqnarray}
%======================================%
where $n(y)$ and $\psi(y)$ are functions of $y$, and $p$, $l$ and
$y_{\pm}$ ($y_-<y_+$) are constants. Note that $t$, $y$ and $x^i$ are
dimensionless and that $n(y)$ and $l$ have the dimension of length.

\subsection{Equations}

With this ansatz, the Einstein equation in the bulk reduces to 
%============< EQUATION >==============%
%
\begin{eqnarray}
 3\alpha^2 p - 1 & = & 0, \nonumber\\
 (\ln n - p\psi)' & = & 0, \nonumber\\
 \left(\psi'\right)^2 - 1 
  \pm \frac{6\alpha^4}{l^2(4-3\alpha^2)}n^2e^{-2\psi} & = & 0, 
\end{eqnarray}
%======================================%
where the plus and minus signs correspond to $V_0>0$ and $V_0<0$,
respectively. Hence, we obtain 
%============< EQUATION >==============%
%
\begin{eqnarray}
 p & = &  \frac{1}{3\alpha^2}, \nonumber\\
 n & = & n_0 e^{p\psi},
\end{eqnarray}
%======================================%
where $n_0$ is an arbitrary constant and $\psi$ is a solution to the
following equation. 
%============< EQUATION >==============%
%
\begin{equation}
 \left(\psi'\right)^2 - 1 \pm e^{2(p-1)\psi} = 0, 
   \label{eqn:master-eq}
\end{equation}
%======================================%
where we have set 
%============< EQUATION >==============%
%
\begin{equation}
 l = \sqrt{\frac{6}{|4-3\alpha^2|}}\alpha^2n_0,
\end{equation}
%======================================%
and the plus and minus signs correspond to $(4-3\alpha^2)V_0>0$ and
$(4-3\alpha^2)V_0<0$, respectively, and we have assumed that
$\alpha^2\ne 4/3$.

The Israel junction condition (\ref{eqn:junction-cond}) is written as 
%============< EQUATION >==============%
%
\begin{equation}
 \left.e^{-(p-1)\psi}\psi'\right|_{y=y_{\pm}} = \pm\gamma_{\pm}, \quad
  \gamma_{\pm} \equiv 
  \frac{\alpha^2\kappa_5\lambda_0^{\pm}n_0}{2\sqrt{|V_0|}l}
  =   \frac{\kappa_5\lambda_0^{\pm}}{2\sqrt{|V_0|}}
  \sqrt{\frac{|4-3\alpha^2|}{6}}.
  \label{eqn:bc-master-eq}
\end{equation}
%======================================%
The scalar field matching condition (\ref{eqn:matching-cond}) is
actually the same as the above junction condition and, thus, does not
provide any independent boundary conditions.

The physically relevant quantities are the power index
$p=\frac{1}{3\alpha^2}$ of the expansion of the branes, the warp factor 
$W\equiv n(y_+)/n(y_-)$, and the ratio $v_{\pm}$ of the inter-brane
distance $L=t\int_{y_-}^{y_+}n(y)dy$ to the proper time
$\tau_{\pm}=n(y_{\pm})t$ on each brane at $y=y_{\pm}$. These are given
by 
%============< EQUATION >==============%
%
\begin{equation}
 W \equiv \frac{n(y_+)}{n(y_-)} = \frac{\tau_+}{\tau_-}
  = e^{p[\psi(y_+)-\psi(y_-)]}, \quad
 v_{\pm} \equiv \frac{L}{\tau_{\pm}} 
  = e^{-p\psi(y_{\pm})}\int_{y_-}^{y_+}e^{p\psi(y)}dy. 
\end{equation}
%======================================%

\subsection{Solutions}

For $(4-3\alpha^2)V_0<0$, the solution to the bulk equation is
%============< EQUATION >==============%
%
\begin{equation}
 e^{-(p-1)\psi} = 
  \left\{\begin{array}{l}
   \sinh\left[|p-1|(y-y_0)\right] \quad (\mbox{for } y_0<y)\\
    \sinh\left[|p-1|(y_0-y)\right] \quad (\mbox{for } y<y_0)
         \end{array}\right. ,
\end{equation}
%======================================%
where $y_0$ is a constant. Note that $\psi'$ diverges at $y=y_0$. Hence,
$y_0$ must not be between the two branes. For this solution, the
junction condition reduces to 
%============< EQUATION >==============%
%
\begin{equation}
 \left\{\begin{array}{l}
  \cosh[(p-1)(y_{\pm}-y_0)] = \mp\gamma_{\pm} \quad
	  (\mbox{for } y_0<y_-<y_+)\\
	 \cosh[(p-1)(y_{\pm}-y_0)] = \pm\gamma_{\pm} \quad
	  (\mbox{for } y_-<y_+<y_0)
	\end{array}\right. .
\end{equation}
%======================================%
Hence, if and only if $1<\gamma_-<-\gamma_+$ 
or $1<\gamma_+<-\gamma_-$, the junction condition uniquely
determines $y_{\pm}-y_0$. The physically relevant quantities $W$ and
$v_{\pm}$ are given by 
%============< EQUATION >==============%
%
\begin{equation}
 W = \left(\frac{\gamma_-^2-1}{\gamma_+^2-1}\right)^{\frac{p}{2(p-1)}},
  \quad
  v_{\pm} = (\gamma_{\pm}^2-1)^{\frac{p}{2(p-1)}}
  \int_{y_-}^{y_+}\frac{dy}
  {\left[\sinh|(p-1)(y-y_0)|\right]^{\frac{p}{p-1}}}.
  \label{eqn:W-vpm-V0negative}
\end{equation}
%======================================%
Not only $W$ but also $v_{\pm}$ are uniquely determined by $p$ (or
equivalently $|\alpha|$) and $\gamma_{\pm}$. For example, when $p=2$, 
%============< EQUATION >==============%
%
\begin{equation}
  \sqrt{v_+v_-} = 
   |\gamma_+|\sqrt{\gamma_-^2-1} + |\gamma_-|\sqrt{\gamma_+^2-1}.
\end{equation}
%======================================%

For $(4-3\alpha^2)V_0>0$, the solution to the bulk equation is
%============< EQUATION >==============%
%
\begin{equation}
 e^{-(p-1)\psi} = \cosh\left[(p-1)(y-y_0)\right],
\end{equation}
%======================================%
where $y_0$ is a constant. For this solution, the junction condition 
reduces to
%============< EQUATION >==============%
%
\begin{equation}
 \sinh[(p-1)(y_{\pm}-y_0)] = \mp\gamma_{\pm}.
\end{equation}
%======================================%
Hence, if and only if $\gamma_-+\gamma_+<0$ then the junction condition
uniquely determines $y_{\pm}-y_0$. The physically relevant quantities
$W$ and $v_{\pm}$ are given by 
%============< EQUATION >==============%
%
\begin{equation}
 W = \left(\frac{\gamma_-^2+1}{\gamma_+^2+1}\right)^{\frac{p}{2(p-1)}},
  \quad
  v_{\pm} = (\gamma_{\pm}^2+1)^{\frac{p}{2(p-1)}}
  \int_{y_-}^{y_+}\frac{dy}
  {\left[\cosh|(p-1)(y-y_0)|\right]^{\frac{p}{p-1}}}.
\end{equation}
%======================================%
Again, not only $W$ but also $v_{\pm}$ are uniquely determined by $p$
(or equivalently $|\alpha|$) and $\gamma_{\pm}$. For example, when
$p=2$, 
%============< EQUATION >==============%
%
\begin{equation}
  \sqrt{v_+v_-} = 
   -\gamma_+\sqrt{\gamma_-^2+1} - \gamma_-\sqrt{\gamma_+^2+1}.
\end{equation}
%======================================%

%======================================%
%<<<<<<<<<<<<  SECTION V >>>>>>>>>>>>>>%
%======================================%

\section{Stability against linear perturbations}

Now let us argue that the expanding scaling solution is a future
attractor. For this purpose we investigate linear perturbations around
the scaling solution and show stability. The linear perturbations
analyzed in this section include all important physical effects such as
boundary conditions on the branes, fluctuations of brane positions and
scalar waves in the bulk. The bulk scalar waves generated by
fluctuations of brane positions threaten to destabilize the expanding
scaling solution, but we shall explicitly see that this is not the case
and that the scaling solution is stable. In the following, for
simplicity we consider the case where $\psi'(y)\ne 0$ for 
$y_-\leq y\leq y_+$.

Let us consider linear perturbations around the scaling solution found
in the previous section. 
%============< EQUATION >==============%
%
\begin{eqnarray}
 ds^2 & = & -n(t,y)^2dt^2 + a(t,y)^2\delta_{ij}dx^idx^j + b(t,y)^2dy^2
  + 2\epsilon(h_{ty}dtdy + h_{ti}dtdx^i + h_{yi}dydx^i),
  \nonumber\\
 n(t,y) & = & n_0 e^{p\psi(y)}
  \left[1+\epsilon\delta n(t,y)\right],
  \nonumber\\
 a(t,y) & = & t^p n_0 e^{p\psi(y)}
  \left[1+\epsilon\delta a(t,y)\right],
  \nonumber\\
 b(t,y) & = & t n_0 e^{p\psi(y)}
  \left[1+\epsilon\delta b(t,y)\right],
  \nonumber\\
 \alpha\kappa_5\phi(t,y) & = & 
  \ln \left[\kappa_5\sqrt{|V_0|}\ lt\right]
  + \psi(y)+\epsilon\delta\psi(t,y),
\end{eqnarray}
%======================================%
where $p$ and $\psi(y)$ are given in the previous section. From the
symmetry, we set 
%============< EQUATION >==============%
%
\begin{equation}
 h_{ty} = h_{ti} = h_{yi} = 0. 
\end{equation}
%======================================%
By introducing a new time coordinate
%============< EQUATION >==============%
%
\begin{equation}
 \tau \equiv \ln t,\quad (-\infty < \tau < \infty),
\end{equation}
%======================================%
it is shown that all coefficients in all relevant equations (i.e., the
Einstein equation, the Israel junction condition and the scalar field
matching condition) linearized with respect to $\epsilon$ are
independent of $\tau$. Thus, it is convenient to Fourier expand the
perturbations as
%============< EQUATION >==============%
%
\begin{eqnarray}
 \delta n(t,y) & = & \delta\bar{n}(y)e^{-i\omega\tau},
  \nonumber\\
 \delta a(t,y) & = & \delta\bar{a}(y)e^{-i\omega\tau}, 
  \nonumber\\
 \delta b(t,y) & = & \delta\bar{b}(y)e^{-i\omega\tau}, 
  \nonumber\\
 \delta\psi(t,y) & = & \delta\bar{\psi}(y)e^{-i\omega\tau}. 
\end{eqnarray}
%======================================%

\subsection{Gauge conditions}

The set of variables ($h_{ty,ti,yi}$, $\delta n$, $\delta a$, 
$\delta b$, $\delta\psi$) includes not only physical degrees of freedom
but also gauge degrees of freedom. The infinitesimal gauge
transformation is 
%============< EQUATION >==============%
%
\begin{eqnarray}
 \delta g_{\mu\nu} & \to & \delta g_{\mu\nu} 
  - \epsilon\nabla_{\mu}\xi_{\nu}- \epsilon\nabla_{\nu}\xi_{\mu}, \nonumber\\
 \delta\phi & \to & \delta\phi 
  - \epsilon\xi^{\mu}\nabla_{\mu}\phi^{(0)},
\end{eqnarray}
%======================================%
where $\nabla$ is the covariant derivative compatible with the
background metric, $\phi^{(0)}$ is the background of $\phi$ and
$\epsilon\xi_{\mu}$ is an infinitesimal vector representing the gauge
degrees of freedom. Hence, each component transforms as 
%============< EQUATION >==============%
%
\begin{eqnarray}
 \delta\bar{n} & \to & \delta\bar{n} + \frac{1}{n_0^2}
  \left[(1-i\omega)\bar{\xi}_t(y)-p\psi'(y)\bar{\xi}_y(y)\right],
  \nonumber\\
 \delta\bar{a} & \to & \delta\bar{a} + \frac{1}{n_0^2}
  \left[p\bar{\xi}_t(y)-p\psi'(y)\bar{\xi}_y(y)
   -\bar{\xi}_{\parallel}(y)\right],\nonumber\\
 \delta\bar{b} & \to & \delta\bar{b} + \frac{1}{n_0^2}
  \left[\bar{\xi}_t(y)-p\psi'(y)\bar{\xi}_y(y)-\bar{\xi}'_y(y)\right],
  \nonumber\\
 h_{ty} & \to & h_{ty} + te^{2p\psi(y)}
  \left[i\omega\bar{\xi}_y(y)-\bar{\xi}_t'(y)\right], \nonumber\\
 h_{ti} & \to & h_{ti} 
  + i\omega t^{2p-1}e^{2p\psi(y)}\bar{\xi}_{\parallel}(y)
  \delta_{ij}(x^j-x_0^j), \nonumber\\
 h_{yi} & \to & h_{yi} 
  - t^{2p}e^{2p\psi(y)}\bar{\xi}_{\parallel}'(y)\delta_{ij}(x^j-x_0^j),
  \nonumber\\
 \delta\bar{\psi} & \to & \delta\bar{\psi} 
  + \frac{1}{n_0^2}[\bar{\xi}_t(y)-\psi'(y)\bar{\xi}_y(y)],
\end{eqnarray}
%======================================%
where we have Fourier expanded $\xi_{\mu}$ as
%============< EQUATION >==============%
%
\begin{eqnarray}
 \xi_t & = & te^{2p\psi(y)}\bar{\xi}_t(y)e^{-i\omega\tau}, \nonumber\\
 \xi_y & = & t^2e^{2p\psi(y)}
  \bar{\xi}_y(y)e^{-i\omega\tau}, \nonumber\\
 \xi_i & = & t^{2p}e^{2p\psi(y)}\bar{\xi}_{\parallel}(y)e^{-i\omega\tau}
  \delta_{ij}(x^j-x_0^j).
\end{eqnarray}
%======================================%
Here, $x_0^j$ ($j=1,2,3$) are constants. The symmetry assumption 
$h_{ty}=h_{ti}=h_{yi}=0$ is consistent with the gauge transformation if
and only if 
%============< EQUATION >==============%
%
\begin{equation}
  i\omega\bar{\xi}_y(y)-\bar{\xi}_t'(y) 
   = \omega \bar{\xi}_{\parallel}(y)
   = \bar{\xi}_{\parallel}'(y) = 0. 
\end{equation}
%======================================%

By using the gauge freedom, we can set 
%============< EQUATION >==============%
%
\begin{equation}
 \delta\bar{\psi} = 0.  \label{eqn:gauge-cond}
\end{equation}
%======================================%
However, this condition does not fix the gauge freedom completely. So,
let us investigate the residual gauge freedom. 

(i) For $\omega\ne 0$ and $-i\omega\ne p-1$, the residual gauge freedom
is
%============< EQUATION >==============%
%
\begin{eqnarray}
 \delta\bar{n} & \to & \delta\bar{n} -\frac{i\omega+p-1}{n_0^2}\bar{\xi}(y),
  \nonumber\\
 \delta\bar{a} & \to & \delta\bar{a},\nonumber\\
 \delta\bar{b} & \to & \delta\bar{b}
  -\frac{i\omega+p-1}{[n_0\psi']^2}\bar{\xi}(y),
  \label{eqn:residual-gauge-1}
\end{eqnarray}
%======================================%
where $\bar{\xi}$ ($=\bar{\xi}_t$) is an arbitrary solution of
%============< EQUATION >==============%
%
\begin{equation}
 \psi'\bar{\xi}' = i\omega \bar{\xi}. 
\end{equation}
%======================================%
Here, we have used the background equation (\ref{eqn:master-eq}) and its
derivative.

(ii) For $\omega\ne 0$ and $-i\omega=p-1$, there is no residual gauge
freedom.

(iii) For $\omega= 0$ and $p\ne 1$, the residual gauge freedom is
%============< EQUATION >==============%
%
\begin{eqnarray}
 \delta\bar{n} & \to & \delta\bar{n} + \delta\bar{n}_0,\nonumber\\
 \delta\bar{a} & \to & \delta\bar{a} + \delta\bar{a}_0,\nonumber\\
 \delta\bar{b} & \to & \delta\bar{b} 
  + \frac{\delta\bar{n}_0}{[\psi'(y)]^2}
  \label{eqn:residual-gauge-3}
\end{eqnarray}
%======================================%
where $\delta n_0$ and $\delta a_0$ are constants.

(iv) For $\omega=0$ and $p=1$, the residual gauge freedom is
%============< EQUATION >==============%
%
\begin{equation}
 \delta\bar{a} \to \delta\bar{a} + \delta\bar{a}_0,
  \label{eqn:residual-gauge-4}
\end{equation}
%======================================%
where $\delta a_0$ is a constant.

\subsection{Linearized equations}

By using $\tau$ as a time coordinate, the Einstein equation linearized
with respect to $\epsilon$ becomes relatively simple in the sense that
there is no explicit $\tau$ dependence in any coefficients. The Israel
junction condition and the scalar field matching condition also share
this property. This is the reason why we have Fourier-expanded the
linear quantities ($\delta n$ $\delta a$, $\delta b$, $\delta\psi$) with
respect to $\tau$. Hence, what we have is just an eigenvalue problem in 
one-dimension.

The ($1$d) bulk equations are 
%============< EQUATION >==============%
%
\begin{eqnarray}
 \psi'\delta\bar{n}' & = & 
  -\frac{\omega^2+(p-1)i\omega}{p}\delta\bar{a} 
  +(p-1)\left(\psi'\right)^2\delta\bar{b}
  + \left[i\omega-(p-1)\right]\delta\bar{n}, \nonumber\\
 \psi'\delta\bar{a}' & = & 
  - i\omega \delta\bar{a} 
  + p\left(\psi'\right)^2\delta\bar{b}
  - p\delta\bar{n}, \nonumber\\
 \psi'\delta\bar{b}' & = & 
  - \frac{\omega^2+(p-1)i\omega}{p}\delta\bar{a} 
  + \left[i\omega + 2(p-1) - (4p-1)\left(\psi'\right)^2\right]\delta\bar{b}
  + (2p+1)\delta\bar{n},  \label{eqn:1st-order-eq}
\end{eqnarray}
%======================================%
where the plus and minus signs correspond to $(4p-1)V_0>0$ and
$(4p-1)V_0<0$, respectively.

Since the gauge freedom has been fixed only up to the residual gauge
freedom, the brane position is not at $y=y_{\pm}$ any more. Hence, we
need to consider the perturbed position of the brane: 
%============< EQUATION >==============%
%
\begin{equation}
 y = Y_{\pm}(\tau) \equiv
  y_{\pm} + \epsilon\delta\bar{y}_{\pm}e^{-i\omega\tau}. 
  \label{eqn:perturbed-position}
\end{equation}
%======================================%
For the fluctuating position of the brane, the formulas
(\ref{eqn:junction-cond}) and (\ref{eqn:matching-cond}) cannot be
used. As explained in Appendix~\ref{app:boundary-cond}, the general
boundary condition (\ref{eqn:general-boundary-cond}) gives the following
boundary condition for perturbations: 
%============< EQUATION >==============%
%
\begin{eqnarray}
 \delta\bar{n}' & = & i\omega\psi'\delta\bar{b}, \nonumber\\
 \delta\bar{a}' & = & 0, \nonumber\\
 (p-1+i\omega) \delta\bar{y}_{\pm} & = & -\psi'\delta\bar{b}. 
  \label{eqn:1st-order-bc}
\end{eqnarray}
%======================================%
Note that these boundary conditions are imposed on $y=y_{\pm}$ (not
$Y_{\pm}(y)$).

\subsubsection{Second-order equation for $\delta\bar{a}$}

From the set of equations (\ref{eqn:1st-order-eq}) and the background
equation (\ref{eqn:master-eq}), we can show that $\delta\bar{a}$
satisfies the following second-order ordinary differential equation: 
%============< EQUATION >==============%
%
\begin{equation}
 \delta\bar{a}'' + 3p\psi'\delta\bar{a}' + \omega(\omega+3ip)\delta\bar{a}
  = 0. \label{eqn:eq-for-delta-a}
\end{equation}
%======================================%
Now let us show that $\Im\omega<0$ unless $\delta\bar{a}\equiv 0$. By
using the equation (\ref{eqn:eq-for-delta-a}) with the boundary
condition $\delta\bar{a}'=0$ at $y=y_{\pm}$, it is easy to show that 
%============< EQUATION >==============%
%
\begin{equation}
 \omega(\omega+3ip)\int_{y_-}^{y_+}dy e^{3p\psi}
  \left|\delta\bar{a}\right|^2
  = \int_{y_-}^{y_+}dy e^{3p\psi}
  \left|\delta\bar{a}'\right|^2. 
\end{equation}
%======================================%
Hence, $\omega(\omega+3ip)$ is real and non-negative unless
$\delta\bar{a}\equiv 0$. The eigenvalue $\omega(\omega+3ip)$ can vanish
only if $\delta\bar{a}'=0$. Thus, 
%============< EQUATION >==============%
%
\begin{equation}
 -3p\leq\Im\omega<0,
\end{equation}
%======================================%
or 
%============< EQUATION >==============%
%
\begin{equation}
 \omega = 0, \quad \delta\bar{a}'=0. 
\end{equation}
%======================================%
In the latter case, $\delta\bar{a}$ can be set to zero by using the
residual gauge freedom (\ref{eqn:residual-gauge-3}) or
(\ref{eqn:residual-gauge-4}). Therefore, we have shown that 
 $-3p\leq\Im\omega<0$ or $\delta\bar{a}\equiv 0$.

\subsubsection{Solution with $\delta\bar{a}=0$}

When $\delta\bar{a}=0$, the set of first-order equations
(\ref{eqn:1st-order-eq}) reduces to 
%============< EQUATION >==============%
%
\begin{eqnarray}
 \psi'\delta\bar{n}' & = & i\omega\delta\bar{n}, \nonumber\\
 (\psi')^2\delta\bar{b} & = & \delta\bar{n}. \label{eqn:da=0-solution}
\end{eqnarray}
%======================================%
For $-i\omega\ne p-1$, the boundary condition (\ref{eqn:1st-order-bc})
is automatically satisfied, but any solutions to
(\ref{eqn:da=0-solution}) with $\delta\bar{a}=0$ is pure gauge because 
of the residual gauge freedom (\ref{eqn:residual-gauge-1}) or
(\ref{eqn:residual-gauge-3}).  
On the other hand, for $-i\omega=p-1$, there is no residual gauge
freedom, but the third boundary condition in (\ref{eqn:1st-order-bc})
becomes $\delta\bar{b}=0$ at $y=y_{\pm}$ and, thus,
(\ref{eqn:da=0-solution}) implies that $\delta\bar{b}=\delta\bar{n}=0$.
Therefore, we have shown that there is no non-vanishing physical
solution with $\delta\bar{a}=0$. 

\subsubsection{Stability}

In summary, we have found that there is no unstable mode with 
$\delta\bar{a}\ne 0$ and that there is no physical mode with
$\delta\bar{a}=0$. Thus, the scaling solution is stable against linear
perturbations.

%======================================%
%<<<<<<<<<<<<< SECTION VI >>>>>>>>>>>>>%
%======================================%

\section{Towards a non-linear stability analysis}

Following the linear perturbation  stability analysis, we need to
establish the stability at the non-linear level. In general this is very complicated.
However, we can investigate some of the non-linear effects by considering a subclass
of models with a particular ansatz. This will give us an indication of the qualitative
features of the
non-linear dynamics. We shall perform a more comprehensive analysis at a later time.

The ansatz we consider is
%============< EQUATION >==============%
%
\begin{eqnarray}
 ds^2 & = & n_0^2 e^{2p\psi(y)}
  \left[-N(t)^2dt^2 + A(t)^2\delta_{ij}dx^idx^j + B(t)^2dy^2\right], 
  \nonumber\\
 \tilde{\alpha}\kappa_5\phi & = & \Theta(t) + \psi(y),
\end{eqnarray}
%======================================%
where $n_0$ is a constant with the dimension of length. The scaling
solution is, of course, included within this ansatz. The ($ty$)-component of
Einstein equation leads to
%============< EQUATION >==============%
%
\begin{equation}
 \Theta(t) = \ln[B(t)] + \Theta_0,
\end{equation}
%======================================%
where $\Theta_0$ is an arbitrary constant. Since $\Theta_0$ can be
absorbed by a redefinition of $\psi(y)$, for convenience we can choose
$\Theta_0$  as 
%============< EQUATION >==============%
%
\begin{equation}
 e^{2\Theta_0} = \frac{2\kappa_5^2n_0^2|V_0|}{3p|4p-1|},
\end{equation}
%======================================%
whence the field equation reduces to 
%============< EQUATION >==============%
%
\begin{eqnarray}
 (1+2p)\frac{\ddot{A}}{A} - 2(p-1)\frac{\dot{A}^2}{A^2} 
  - \left[(4p-1)\frac{\dot{B}}{B}+(2p+1)\frac{\dot{N}}{N}\right]\frac{\dot{A}}{A}
  +3p^2\frac{\dot{B}^2}{B^2} & = & 0,\nonumber\\
 (1+2p)\frac{\ddot{B}}{B} +3p\frac{\dot{B}^2}{B^2} 
  + \left[3(2p-1)\frac{\dot{A}}{A}-(2p+1)\frac{\dot{N}}{N}\right]\frac{\dot{B}}{B}
  -6\frac{\dot{A}^2}{A^2} & = & 0, \nonumber\\
 \left(\psi'\right)^2 \pm e^{2(p-1)\psi} - \frac{1}{p(2p+1)N^2}
  \left[ 2\frac{\dot{A}^2B^2}{A^2} + 2\frac{\dot{A}B\dot{B}}{A}-p\dot{B}^2\right]
  & = & 0, \label{eqn:nonlinear-dynamical-eq}
\end{eqnarray}
%======================================%
where the plus and minus signs in the last equation are for
$(4p-1)V_0>0$ and $(4p-1)V_0<0$, respectively. Hence, the model is
{\it separable}.

Hereafter, we set $N(t)$ to a constant by choice of time
coordinate. Defining 
%============< EQUATION >==============%
%
\begin{equation}
 a = \frac{\dot{A}}{A}, \quad b = \frac{\dot{B}}{B},
\end{equation}
%======================================%
the last equation in (\ref{eqn:nonlinear-dynamical-eq}) becomes
%============< EQUATION >==============%
%
\begin{equation}
 c^2_0 B^{-2} = 2a^2 + 2ab - pb^2, \label{eqn:(C)}
\end{equation}
%======================================%
where $c^2_0$ is effectively the rescaled separation constant, and 
the evolution equations are then
%============< EQUATION >==============%
%
\begin{eqnarray}
 (1+2p)\dot{a} & = & -3a^2 + (4p-1) ab -3p^2b^2, \label{eqn:(A)}\\
 (1+2p)\dot{b} & = &  6a^2 - 3(2p-1) ab -(1+5p)b^2. \label{eqn:(B)}
\end{eqnarray}
%======================================%
Differentiating equation (\ref{eqn:(C)}) and using equations
(\ref{eqn:(A)}) and (\ref{eqn:(B)}) we get an expression which is
satisfied identically, so that (\ref{eqn:(C)}) is a constraint (that 
propagates along the solution curves).

We first note that since the system (\ref{eqn:(A)}), (\ref{eqn:(B)}) is 
homogeneous, we can define
%============< EQUATION >==============%
%
\begin{equation}
 x = a/b, \quad \overline{b} = ln b,
\end{equation}
%======================================%
and the system reduces to a single ordinary differential equation:
%============< EQUATION >==============%
%
\begin{equation}
 \frac{dx}{d\overline{b}} = \frac{6(x-p) (x+\frac{1}{2}
  (1+\sqrt{1+2p}))(x+\frac{1}{2} (1 - \sqrt{1+2p}))}
  {(1+5p) + 3(2p-1)x - 6x^2}
\end{equation}
%======================================%
which can be integrated to obtain
%============< EQUATION >==============%
%
\begin{equation}
 b=b^2_0 
  (x-p)^{\tilde{\alpha}}
  (x + \frac{1}{2}(1+\sqrt{1+2p}))^{\tilde{\beta}}
  (x+\frac{1}{2}(1-\sqrt{1+2p}))^{\tilde{\gamma}}, \label{eqn:dag}
\end{equation}
%======================================%
where $b^2_0$ is an integration constant and
%============< EQUATION >==============%
%
\begin{eqnarray}
 \tilde{\alpha} & \equiv & \frac{1}{3p}, \nonumber\\
 \tilde{\beta} & \equiv & \frac{-3p-1+\sqrt{2p+1}}{6p}
  = -\frac{(3\sqrt{2p+1}+1)(\sqrt{2p+1}-1)}{12p}
  = -1 + \frac{(3\sqrt{2p+1}+5)(\sqrt{2p+1}-1)}{12p},\nonumber\\
 \tilde{\gamma} & \equiv & \frac{-3p-1-\sqrt{2p+1}}{6p}
  = -\frac{(3\sqrt{2p+1}-1)(\sqrt{2p+1}+1)}{12p}
  = -1 + \frac{(3\sqrt{2p+1}-5)(\sqrt{2p+1}+1)}{12p}. \label{eqn:star}
\end{eqnarray}
%======================================%
Note that
%============< EQUATION >==============%
%
\begin{eqnarray}
 \tilde{\alpha}+\tilde{\beta}+\tilde{\gamma} & = & -1, \nonumber\\
 \tilde{\alpha} & > & 0, \nonumber\\
 -1< \tilde{\beta} & < & 0, \nonumber\\
 \tilde{\gamma} & < & 0
\end{eqnarray}
%======================================%
for any positive $p=1/(3\alpha^2)$, and $\tilde{\gamma}>-1$ if $p>8/9$.

We can analyze the asymptotic form of (\ref{eqn:dag}), but 
perhaps a better way to show the late-time stability
of the scaling solution is as follows.  We rewrite equation
(\ref{eqn:(C)}) as 
%============< EQUATION >==============%
%
\begin{equation}
 1 - \frac{(2p+1)b^2}{4(a+\frac{1}{2} b)^2}
  = \frac{c^2_0 B^{-2}}{2(a+\frac{1}{2}b)^2}. 
\end{equation}
%======================================%
Defining
%============< EQUATION >==============%
%
\begin{equation}
 z = \frac{b\sqrt{1+2p}}{2a+b}
  = \frac{\sqrt{1+2p}}{1+2x}, \label{eqn:(D1)}
\end{equation}
%======================================%
we see that $z$ is bounded; $1-z^2 \geq 0$.  Using (\ref{eqn:dag}), the
evolution equations (\ref{eqn:(A)}) and (\ref{eqn:(B)}) then become a
single ordinary differential equation for $z$:  
%============< EQUATION >==============%
%
\begin{equation}
 \dot{z} = \frac{3b_0^2}{2^{2\tilde{\beta}}\sqrt{1+2p}}
  (1-\sqrt{1+2p}\; z)^{\tilde{\alpha}+1} (1+z)^{\tilde{\beta}+1} 
  (1-z)^{\tilde{\gamma} +1}. \label{eqn:ddag}
\end{equation}
%======================================%
Since $z^2\leq 1$, this constitutes a one-dimensional non-linear
dynamical system.

The exponent $(\tilde{\alpha} +1)$ is positive definite, so that
%============< EQUATION >==============%
%
\begin{equation}
 z_s =  \frac{1}{\sqrt{1+2p}}      \label{eqn:(D2)}
\end{equation}
%======================================%
(where $0 < z_s < 1$), is always an equilibrium point of
(\ref{eqn:ddag}). Depending on the signs of the exponents
$(\tilde{\beta}+1)$ and $(\tilde{\gamma} +1)$, $z_\pm= \pm 1$ are also
equilibrium points ($z_-$ is always an equilibrium point, while
$z_+$ is an equilibrium point for $p> 8/9$). 
  
The solution of (\ref{eqn:ddag}) close to $z_s$ is given by 
%============< EQUATION >==============%
%
\begin{equation}
 z = z_s-   z^2_0 (t+t_0)^{-1/\tilde{\alpha}},
\end{equation}
%======================================%
so that $z \to z_s$ as $t\to \infty$ since $\tilde{\alpha}>0$. Hence,
$z_s$ is a local sink. Since $z$ is bounded, $z_s$ is a global
attractor in the physical phase space. Indeed, since $\tilde{\beta}<0$ 
and $\tilde{\gamma}<0$, $z_{\pm}$ act as local sources and all physical
solutions evolve from one of $z_+$ and $z_ -$ to $z_s$.

From equations (\ref{eqn:(D1)}) and (\ref{eqn:(D2)}), the global
attractor has 
%============< EQUATION >==============%
%
\begin{equation}
 \frac{a}{b} \; (=x) = p
\end{equation}
%======================================%
so that equation (\ref{eqn:(C)}) yields 
%============< EQUATION >==============%
%
\begin{equation}
 \dot{B}^2 = \frac{c_0^2}{p(2p+1)}  = \mbox{ const.}
\end{equation}
%======================================%
and hence after a time translation 
%============< EQUATION >==============%
%
\begin{equation}
 B = B_0 t, \quad A = A_0 t^p,
  \label{eqn:attractor}
\end{equation}
%======================================%
so that the global attractor is the scaling solution.

To connect with earlier work, the approximate solution of
(\ref{eqn:(A)}) and (\ref{eqn:(B)}) at late times (i.e., the linearized 
solution around the attractor (\ref{eqn:attractor})) is ($N\equiv 1$): 
%============< EQUATION >==============%
%
\begin{eqnarray}
 A & = & A_0t^p\left[ 1 + 3pc_1t^{-1} + c_2t^{-3p}\right], 
  \nonumber\\
 B & = & B_0t\left[ 1 - \frac{c_1}{p}t^{-1} - \frac{c_2}{p}t^{-3p}\right], 
\end{eqnarray}
%======================================%
where the $c_i$ are arbitrary constants (and the constants are subject
to the constraint (\ref{eqn:(C)}); e.g., $c_0^2=p(1+2p)B_0^2$). From
this we immediately see the decaying modes and the local stability of
the attractor.

Consequently, we have an approximate bulk ($5$d) solution at late
times. The $5$d Ricci tensor is given by its attractor values of order
$O(t^{-1})$ and $O(t^{-3p})$ (leading to $\rho_{kin}\sim t^{-2}$), and
the $5$d conformal tensor (which is zero for the exact attractor
solution) can be evaluated to leading order. We can therefore calculate
the irreducible decompositions ${\cal U}$, ${\cal Q}_\mu$, and ${\cal
P}_{\mu\nu}$ of the projected bulk Weyl tensor ${\cal
E}_{\mu\nu}$~\cite{SMS,Maartens}, from which we find that 
%============< EQUATION >==============%
%
\begin{equation}
 {\cal Q}_{\mu} = 0, {\cal P}_{\mu\nu}=0
\end{equation}
%======================================%
and
%============< EQUATION >==============%
%
\begin{equation}
 {\cal U} \sim \left[ O(t^{-3p}) + O(t^{-2})\right]\rho_{kin}
  \ll \rho_{kin}. 
\end{equation}
%======================================%

%======================================%
%<<<<<<<<<<<< SECTION VII >>>>>>>>>>>>>%
%======================================%
  
\section{Discussion}

% Scaling solution

We have investigated a class of scale-invariant effective theories of
$5$d brane-world cosmology with a bulk scalar between two
end-of-the-world branes. As a special case, this class includes the $5$d
reduction of Horava-Witten theory. We have found an exact, analytic
scaling solution in which the scale factor of our $4$d universe and the
inter-brane distance expand as $t^p$ ($p>0$) and $t$,
respectively~\footnote{After
submission of this paper, the authors were informed of
ref.~\cite{Koyama-Takahashi}, in which a closely related solution had 
been found.}. 
It is perhaps worth mentioning that the scaling solution with $p>1$
corresponds to a power-law inflation on our brane. The scaling solution
corresponds to an equilibrium state in which the motion of the branes
does not produce any scalar waves in the bulk. Because of this
remarkable physical property, the scaling solution is expected to be a
future attractor of the system. Indeed, we have presented convincing
evidence for the attractor behavior: the stability of the scaling
solution against general linear perturbations with the $4$d FRW symmetry
and stability within a class of non-linear perturbations.

% Scenario (scaling solution -> radion stabilization)

Based on the attractor behavior of the scaling solution, in
Sec.~\ref{sec:scenario} we proposed a new, self-consistent and dynamical
scenario in the early brane-world universe. First, since the scaling
solution is a future attractor, the system is automatically driven
towards it as far as the effective action at high energy is of the scale
invariant form. Second, as the energy scale becomes sufficiently low
according to the evolution along the scaling solution, the scale
invariance of the effective action should be broken at some point. After
that, the brane-world system deviates from the scaling behavior and
starts to be driven by the radion stabilization. The radion
stabilization guarantees that $4$d gravity on our brane is described by
Einstein theory and the standard cosmology is realized at low energy.

In this scenario, the scaling solution plays a central role: the
attractor behavior of the scaling solution makes it possible to give
well-defined initial conditions both in the bulk and on the brane for
the evolution after the breaking of scale invariance. It is also the
scaling solution that brings the system from high energy to low energy
both in the bulk and on the brane and, thus, triggers the breaking of
the scaling invariance.

% Question I (low energy)

Having a well-defined initial condition given by the scaling solution, a
natural question arises: ``What kind of evolution can we expect
subsequently?'' One interesting possibility is an epoch of inflation
driven by the radion (i.e., the inter-brane distance). As stated above,
after the scale invariance of the effective theory is broken at some
point, the radion stabilization mechanism takes over and starts driving
the brane-world system. Indeed, we have suggested a simple illustrative
action in which the transition from the scale-invariance to the radion
stabilization is smooth. During the transition, the dynamics of the bulk
scalar field and the radion can drive the evolution of our $4$d
universe. Of course, even without the scaling behavior at high energy,
the bulk scalar and the radion can drive the system. A big difference,
however, is that one cannot expect a well-defined, predictable initial
condition in the case without the high-energy scaling behavior since the
evolution from an arbitrary state to the radion stabilization should be
a very violent process and should involve fully non-linear (scalar and
gravitational) waves in the bulk. Our scenario based on the high-energy 
scaling behavior makes the evolution towards the radion stabilization
much smoother and, more importantly, predictable. 
In particular, in our scenario, there may be an inflation in the
transition epoch. This possibility is certainly an interesting subject
for the future work.

% Question  II (high energy)

Another question that might naturally be asked is: ``What is the
beginning of the brane-world before the scaling behavior?'' Actually,
since the attractor behavior of the scaling solution would make the
low-energy evolution of the brane cosmology almost insensitive to the
beginning, this question might be irrelevant for our scenario. 
Nonetheless, it may be still interesting to think about this kind of
question. As already explained, for the scaling solution both the scale
factor of our universe and the inter-brane distance expand. This implies
that both our $4$d universe and the extra dimension were extremely small
at an early epoch. This is a completely trivial statement in our
scenario, but it would not be so trivial if we did not have the
high-energy scaling behavior. Without the high-energy scaling behavior,
the initial value of the inter-brane distance for the evolution towards
the stable value can be either small or large. On the other hand, in our
scenario the initial value of the inter-brane distance should start from
a small value. Is this favorable from the viewpoint of quantum gravity?
Can we consider the smallness of the initial inter-brane distance as an
indication of a brane collision or brane scattering?~\footnote{
It is perhaps interesting to note that the expanding scaling solution we
found is a conformally Kasner geometry and that Kasner-like geometries
were found to be generic collapsing solutions in brane
collisions~\cite{branecode}.}
The last two questions are evidently outside the scope of this paper but
will be interesting future projects.

% Generalization 

It is perhaps worthwhile asking whether we can expect similar scenarios
to work in more general situations and/or theories. When the dynamical 
effects of fields are negligible (eg., the bulk waves), then we might
expect the same qualitative behavior and hence the scenario to work. For
example, we expect predictable, smooth evolution towards the radion 
stabilization to be valid in more general situations whenever a symmetry
at high energy exists, leading to attractor behavior, which is
subsequently broken by radion stabilization at low energy.

% Non-linear analysis, matter fields on brane, higher dimension

As already stressed many times, we have presented convincing evidence
that the scaling solution is a future attractor. Since the attractor
behavior is so essential in our scenario, in future work we shall
further investigate the full non-linear dynamics both analytically and
numerically. We shall also include other matter fields on the brane. 
Since the scale invariant theory is motivated by compactification of
higher dimensional theories (see Sec.~\ref{sec:scenario} and
subsection~\ref{subsec:action}), it is also worth investigating 
higher-dimensional interpretations of the scaling solution. In
particular, it is interesting to generalize the scaling behavior to
higher dimension (greater than $5$) and study the resulting physical 
consequences.

%%%%%%%%%%%%%%%%%%%%%%%%%%%%%%%%%%%%%%%%%%%%%%%%%%%%%%%%%%%%%%%%%%%%
%%%%%%%%%%%%%%%%%%%%%%%%%%%%%%%%%%%%%%%%%%%%%%%%%%%%%%%%%%%%%%%%%%%%
% Acknowledgements
%%%%%%%%%%%%%%%%%%%%%%%%%%%%%%%%%%%%%%%%%%%%%%%%%%%%%%%%%%%%%%%%%%%%
%%%%%%%%%%%%%%%%%%%%%%%%%%%%%%%%%%%%%%%%%%%%%%%%%%%%%%%%%%%%%%%%%%%%
\begin{acknowledgments}
 The authors would like to thank Paolo Creminelli, Andrei Frolov, Lev
 Kofman and Toby Wiseman for their useful comments. SM would like to
 acknowledge the kind hospitality of Dalhousie University where this
 work was initiated during his visit. Part of this work was done during
 SM's visits to University of Toronto, University of Victoria and
 University of Alberta. He would be grateful to Lev Kofman, Werner
 Israel and Valeri Frolov for their warm hospitality and stimulating
 discussions during the visits. SM's research was supported in part by
 NSF grant PHY-0201124. AAC's research was funded by the Natural
 Sciences and Engineering Research Council of Canada. 
\end{acknowledgments}

%%%%%%%%%%%%%%%%%%%%%%%%%%%%%%%%%%%%%%%%%%%%%%%%%%%%%%%%%%%%%%%%%%%%
%%%%%%%%%%%%%%%%%%%%%%%%%%%%%%%%%%%%%%%%%%%%%%%%%%%%%%%%%%%%%%%%%%%%
% Appendix
%%%%%%%%%%%%%%%%%%%%%%%%%%%%%%%%%%%%%%%%%%%%%%%%%%%%%%%%%%%%%%%%%%%%
%%%%%%%%%%%%%%%%%%%%%%%%%%%%%%%%%%%%%%%%%%%%%%%%%%%%%%%%%%%%%%%%%%%%

\appendix

%======================================%
%<<<<<<<< Boundary condition >>>>>>>>>>%
%======================================%

\section{Boundary condition for linear perturbations}
\label{app:boundary-cond}

In this appendix we explain the derivation of the boundary condition
(\ref{eqn:1st-order-bc}) for linear perturbations. For more general
prescription, see ref.~\cite{Mukohyama-junction}. 
For general trajectories of branes the boundary condition is given by 
(\ref{eqn:general-boundary-cond}) at $y=Y_{\pm}(t)$. For
%============< EQUATION >==============%
%
\begin{eqnarray}
 n(t,y) & = & n_0 e^{p\psi(y)}
  \left[1+\epsilon\delta\bar{n}(y)e^{-i\omega\tau}\right],  \nonumber\\
 a(t,y) & = & t^p n_0 e^{p\psi(y)}
  \left[1+\epsilon\delta\bar{a}(y)e^{-i\omega\tau}\right],  \nonumber\\
 b(t,y) & = & t n_0 e^{p\psi(y)}
  \left[1+\epsilon\delta\bar{b}(y)e^{-i\omega\tau}\right],\nonumber\\
 \alpha\kappa_5\phi(t,y) & = & 
  \ln \left[\kappa_5\sqrt{|V_0|}\ lt\right] + \psi(y), \nonumber\\
 Y_{\pm}(\tau) & = &
  y_{\pm} + \epsilon\delta\bar{y}_{\pm}e^{-i\omega\tau},
\end{eqnarray}
%======================================%
the extrinsic curvature components ${\cal K}$ and $\bar{\cal K}$, and
the normal derivative $\partial_{\perp}\phi_{\pm}$ of the scalar field
are expanded as
%============< EQUATION >==============%
%
\begin{eqnarray}
 {\cal K}_{\pm} & = & {\cal K}^{(0)}_{\pm} + \epsilon {\cal K}^{(1)}_{\pm}
  + O(\epsilon^2), \nonumber\\
 \bar{\cal K}_{\pm} & = & \bar{\cal K}^{(0)}_{\pm} 
  + \epsilon \bar{\cal K}^{(1)}_{\pm}e^{-i\omega\tau}
  + O(\epsilon^2), \nonumber\\
 \partial_{\perp}\phi_{\pm} & = &  \partial_{\perp}\phi_{\pm}^{(0)}
  + \epsilon\partial_{\perp}\phi_{\pm}^{(1)}e^{-i\omega\tau}
  + O(\epsilon^2), 
\end{eqnarray}
%======================================%
where
%============< EQUATION >==============%
%
\begin{eqnarray}
 {\cal K}^{(0)}_{\pm} & = & \bar{\cal K}^{(0)}_{\pm}
  =\frac{p\psi'e^{-p\psi}}{n_0t}, \nonumber\\
 \partial_{\perp}\phi_{\pm}^{(0)} & = & 
  \frac{\psi'e^{-p\psi}}{\alpha\kappa_5n_0t},
\end{eqnarray}
%======================================%
and
%============< EQUATION >==============%
%
\begin{eqnarray}
 {\cal K}^{(1)}_{\pm} & = & \frac{e^{-p\psi}}{n_0t}
  \left[ -(i\omega+\omega^2)\delta\bar{y}_{\pm}
   +\delta\bar{n}' -p\psi'\delta\bar{b}\right], \nonumber\\
 \bar{\cal K}^{(1)}_{\pm} & = & \frac{e^{-p\psi}}{n_0t}
  \left[ -ip\omega\delta\bar{y}_{\pm} + \delta\bar{a}'
   - p\psi'\delta\bar{b}\right], \nonumber\\
 \partial_{\perp}\phi_{\pm}^{(1)} & = & 
  \frac{e^{-p\psi}}{\alpha\kappa_5n_0t}
  \left[ -i\omega\delta\bar{y}_{\pm} - \psi'\delta\bar{b}
  \right]. 
\end{eqnarray}
%======================================%

What is important here is that the boundary condition in the form 
(\ref{eqn:general-boundary-cond}) must be imposed on the perturbed
position $y=Y_{\pm}(t)$ of the brane. On the other hand, it is
mathematically convenient to impose a boundary condition on a fixed
position in the coordinate $y$. Hence, let us convert the boundary
condition (\ref{eqn:general-boundary-cond}) at $y=Y_{\pm}(t)$ to a 
boundary condition at $y=y_{\pm}$. The result up to the linear order in
$\epsilon$ is
%============< EQUATION >==============%
%
\begin{eqnarray}
 {\cal K}^{(0)'}_{\pm}\delta\bar{y}_{\pm} + {\cal K}^{(1)}_{\pm}
  & = & \pm\frac{\kappa_5}{6\alpha}(\partial_{\phi_{\pm}}\lambda_{\pm})
  \psi'\delta\bar{y}_{\pm}, 
  \nonumber\\ 
 \bar{\cal K}^{(0)'}_{\pm}\delta\bar{y}_{\pm} + \bar{\cal K}^{(1)}_{\pm}
  & = & \pm\frac{\kappa_5}{6\alpha}(\partial_{\phi_{\pm}}\lambda_{\pm})
  \psi'\delta\bar{y}_{\pm}, 
  \nonumber\\
 (\partial_{\perp}\phi^{(0)}_{\pm})'\delta\bar{y}_{\pm}
  + \partial_{\perp}\phi^{(1)}_{\pm}
  & = & \mp\frac{1}{2\alpha\kappa_5}(\partial_{\phi_{\pm}}^2\lambda_{\pm})
  \psi'\delta\bar{y}_{\pm}. 
 \end{eqnarray}
%======================================%
This form of the boundary condition must be imposed at
$y=y_{\pm}$. Substituting the above expressions for $K^{(i)}_{\pm}$,
$\bar{K}^{(i)}_{\pm}$ and $\partial_{\perp}\phi^{(i)}_{\pm}$ and
simplifying the expressions by using the background equations, we
obtain the boundary condition (\ref{eqn:1st-order-bc}).

%======================================%
%<<<<<<<<<<<< REFERENCES >>>>>>>>>>>>>>%
%======================================%

\end{document}